\documentclass[prd,aps,floats,tabularx,preprintnumbers,twocolumn]{revtex4}
\usepackage{graphicx, epsfig}

\begin{document}

\baselineskip 12 pt

\title{A First Glimpse of String Theory in the Sky?}

\author{Steen Hannestad$^+$, Laura Mersini-Houghton$^\ast$}

\affiliation
{$^+$ Department of Physics, University of Southern Denmark,
Campusvej 55, DK-5230 Odense M, Denmark and
NORDITA, Blegdamsvej 17, DK-2100 Copenhagen, Denmark
\\ $^\ast$ Department of Physics and Astronomy, UNC-Chapel Hill, CB{\#}3255, Phillips Hall,Chapel Hill, NC 27599, USA}

\begin{abstract}
{\small
%%%%%%%

We propose a new method for identifying new physics imprints on present observational data in cosmology whereby signatures of string theory are clearly distinguished from imprints of possible features on the inflaton potential. Our method relies on the cross-correlations spectra of cosmic shear from large scale structure (LSS) with the CMB temperature anisotropies and E-mode polarization, by using the following properties: inflationary cosmology provides only one source term for all CMB spectra and LSS which highly constrains any deviations from the standard predictions; string theory can add new non-inflationary channels to the source of perturbations as well as modify clustering properties at large scales.

Discrepancies in the source terms of correlations and clustering properties provide the evidence for new physics. Models of single-field inflation with a feature are disfavored even with present data. Upcoming WMAP results and future data from weak lensing of LSS will further improve our ability to probe new physics in this manner and could open the first direct window to string theory.}

\end{abstract}

\maketitle

%%%%%%%%%%%%%%%%%%%%%%%%%%5555
\section{Introduction}

There is an emerging picture of the universe obtained from the tremendous progress achieved by precision cosmology. According to this picture we live in a 'weird' universe. A universe which, at present redshifts, $z \simeq 0-1$, and energy scales $H_0 \simeq 10^{-33}eV$, is: dominated by a mysterious component of energy, coined dark energy, driving it into an accelerated expansion \cite{de};and which, contrary to our theoretical expectations based on the inflationary paradigm, has large angle CMB perturbations suppressed \cite{wmap}. The later puzzle is not confirmed yet due to limitations from cosmic variance \cite{cosmicvar}, and the need for a better understanding of the systematic and foreground effects. Hence a lot of work done in investigating this effect \cite{cross, kesden}, by using complementary data to minimize statistical errors, points out that the suppression of CMB power for large angles $\theta$ persists.

In what follows, we take the WMAP CMB measurements \cite{wmap} at large scales as true and consider the power suppression at low multipoles $l$ to be a real physical effect rather than a statistical fluke.

Dark energy domination and CMB power suppression of perturbations at  large angles, both occur, around the same redshift and energy scale $H_0 \simeq 10^{-33}$ eV. They both raise two disturbing questions: Why is their magnitude so small, (the tuning problem), and, why are they occurring now, (the coincidences problem). A consistent theoretical model should simultaneously address both questions, the tuning and the cosmic coincidences, for these observed phenomena. Besides, it should also address why both coincidences occur at the {\it same} energy scale, our present Hubble radius $H_0$, \cite{ale}. It is tempting to speculate that too many coincidences in the present universe may indicate the emergence of a new scale in physics of the order our Hubble scale $H_0 \simeq 10^{-33}$ eV.

No such model exists yet. Perhaps because an understanding of the underlying fundamental theory is required in order to explain the current challenges that theoretical cosmology is facing. As yet string theory is the leading candidate for new physics although it has still not provided us with satisfactory answers to the above issues. A conservative approach would be to question whether we need new physics at all for addressing the current problems. Distinct observational signatures would be the best way for resolving these doubts and for providing direct evidence of new physics if it exists.

Therefore, it is of crucial importance to attempt to distinguish whether what is giving rise to the observed cosmic coincidences are features of inflaton \cite{easther,linde} or imprints of new physics \cite{efstathiou,katie,costa}. This question is the scope of our work here, namely: Can we uniquely identify and discriminate new physics signatures from inflaton features in the presently available observables in the sky?

A positive answer to the above question in favor of new physics, possibly string theory as its leading candidate, would provide the first direct evidence for its existence. This is a difficult task to investigate. Any signatures expected from new physics, e.g. anomalies on the CMB spectra, are very small\cite{tp}. Furthermore, even if they are within the detection limits, it is notoriously difficult to identify whether it is some feature of the unknown inflaton potential\cite{easther} that give rise to the anomalies or whether the anomalies are rooted in new (string) physics. Besides, the origin of dark energy and its relation to the power suppression of low multipoles around redshift $z \simeq 0$ is not yet understood.

In this situation, we need to identify an imprint which arises uniquely from one type of theory but is generically negligible in the other.
We demonstrate in the next section that finding such a unique handle on the observational signatures is possible and then use this handle to discriminate among various models with respect to their imprints on the combined data from CMB spectra, LSS and dark energy. The autocorrelation spectra do not yield any new information on the origin of the observed anomalies since they are sourced by the same term \cite{tp}.

In the standard theory of inflation and physics of the early universe, the temperature perturbations given by the Sachs-Wolfe effect are

\begin{equation}
\label{temp1}
\Theta({\bf n}) = \frac{1}{3} \Phi(r_0,z_0) - 2\int_0^{r_0} \frac{d\Phi}{dr}(r,z(r))dr  \,\,
\end{equation}

where $\Phi(r,z)$ is the gravitational background potential and $r,z$ are the physical comoving distance and redshift respectively. The index $0$ denotes the value at last scattering surface. The gravitational potential $\Phi(r,z)$ is related to the growth factor of structure $G(z)$ through $\Phi(r,z) = (1+z)G(z)\Phi(r,0)$.Denoting the three dimensional matter power spectrum by $P(k)$ then the power spectrum for the CMB temperature anisotropies in a flat universe is

\begin{equation}
\label{temp2}
C_l^{TT} \simeq \int \frac{dk}{k^2} P(k) [\Theta_l (k)]^2 \,
\end{equation}

where $\Theta_l (k)$ is the Fourier transform of $\Theta({\bf n})$.

An expression similar to Eqn.~\ref{temp2} gives the spectrum of the curl free E-mode of polarization of CMB photons, denoted by $C_l^{EE}$ \cite{ma}.

But the same gravitational potential $\Phi(r,z)$ that produces the temperature anisotropies $C_l^{TT}$ also sources the polarization spectra namely the $E, B$ -modes and gravitational lensing of LSS. In fact gravitational lensing can be a powerful tool for mapping the background potential for LSS. The projected spectrum is described by the following potential 

\begin{equation}
\label{lens}
L({\bf n}) = -2\int_0^{r_0} dr \frac{r_0 -r}{r r_o}\Phi(r,z(r)) \,
\end{equation}

and the angular power spectrum of the lensing potential is

\begin{equation}
\label{lens2}
C_l^{LL} \simeq \int \frac{dk}{k^2} P(k) [L_l (k)]^2 \,
\end{equation}

where similarly $L_l (k)$ is the Fourier transform of $L({\bf n})$.
Notice however that although both spectra depend on the same source in the conventional theory, their integral dependence on $\Phi(r,z)$, in the expressions (\ref{temp1}) and (\ref{lens}) are sufficiently different. This fact can be exploited by using the cross correlation spectrum between $T,L$ in identifying disreptancies with the expectations of the standard concordance cosmology.

\begin{equation}
\label{crossshear}
C_l^{TL} \simeq \int \frac{dk}{k^2} P(k)[ \Theta_l (k) L_l (k) ] \,
\end{equation}

A similar expression to Eqn.~(\ref{crossshear}) gives the cross-correlation spectrum $C_l^{TE}$ of T with E-mode.

The role of dark energy on the low multipoles $l$ is to enhance their power through the Integrated Sachs-Wolfe (ISW) effect. The observed suppression of power may indicate that clustering properties at large scales are different from the predictions of the standard theory and such that they compete with the ISW effect.
The correlations reveal whether either of the two spectra deviate, due to new physics, from the standard dependence on $\Phi$ expected by the equations above. Deviations may arise from contributions from noninflationary channels or a fundamental string scale imprinted on $C_l^{TL,TE}$ that may signal a breakdown of the conventional theory.

Below we use the combination of both correlation spectra $C_l^{TE}$ and $C_l^{TL}$  as our handle in identifying the origin of the imprints on observations while avoiding some of the degeneracies among cosmic parameters. Generically, the inflaton field, (in single-field inflation models), is very weakly coupled to other sectors due to the slow roll conditions and adiabaticity requirement. Meanwhile there is no reason why the coupling of the stringy moduli fields to the matter sector should be suppressed. Therefore any features on the inflaton potential and their signature to LSS  and clustering properties at large scales would be very different from the effects from variations of moduli couplings carried over to the matter sector. These moduli couplings could have a large impact onto the clustering properties of large scale structure and the polarization spectrum. For this reason discrepancies in the cross correlations would reveal complimentary information in identifying the origin of the observed signatures.

In Sect.2 we provide the theoretical framework for a generic classification of the possible effects of string and brane-world models on the low-energy universe. There we describe how cross-correlation spectra provide the extra information needed in order to discriminate between a standard inflationary model with a feature on its potential, from the various string-inspired classes of cosmology beyond the concordance model. We focus only on the string models from literature that can give rise to an accelerated expansion of the universe at late time, a criteria imposed by observations, which takes account of the ISW effect.

Data analysis of $C_l^{TT, EE, TE, TL}$ spectra for the string effects in cosmology such as variations of the string coupling constants, dispersion relation for the matter fields arising from their coupling to moduli fields, changes in the strength of gravity, large scale clustering properties and variations of the background gravitational potential $\phi$ due to the higher dimensional nature of gravity, is presented in Sect.3. There we show how the $C_l^{TE}$ spectrum breaks some of the degeneracy of the cosmological parameters while the $C_l^{TL}$ spectrum carries the unique signature for new physics. Results obtained in Sect.3. are discussed in Sect.4.

\section{Theoretical Framework}

Suppose we were to focus solely on the origin of CMB power suppression at large angles $\theta$, and the cosmic coincidence this suppression introduces to cosmology, while putting aside the fine-tuning and coincidence puzzles associated with dark energy.

In this case, a bump on the inflaton potential which can violate the slow roll conditions can be as good of a candidate for giving rise to the observed CMB power suppresion\cite{easther,linde} as any of the more exotic possibilities associated with string theory and brane worlds.

Nevertheless, a feature on the inflaton potential can not address why this suppression occurs only for such specific low 'l' values, that is, only for perturbations modes with wavelength of the order our present Hubble radius $H_0 =10^{-33}eV$. For example for GUT scale inflation a feature has to be introduced in an $ad-hoc$ manner in its potential around the 60th e-folding in order to suppress the power of present low 'l' multipoles.

We will not try to address the issue of the CMB cosmic coincidence\cite{ale} here since the answer may be rooted at a deeper level to an interplay with the dark energy coincidence occurring at the same time and energy scale. However the combination of the two problems, as discussed in \cite{ale}, may be a good hint already that a new and fundamental theory may be at works for these puzzling observations in the late time universe. The double cosmic coincidence provides our motivation for trying to discriminate inflationary signatures versus new physics imprints from observations, specifically with respect to the origin of the observed CMB large scale power suppression.Coupling to the matter sector is our handle for discriminating among the two groups of models: conventional and, new physics. 

The inflaton is very weakly coupled to the matter sector by the constraints of the slow roll and adiabaticity. Any features in its potential around the 60th efolding, $k_{60}$, while they may suppress power of temperature autocorrelations $C_l^{TT}$ at low 'l', they predict an identical behavior  for all other spectra since the source term for seeding all spectra is the same\cite{ma} and can be calculated. Thus their predictable impact on deviations from the conventional large scale structure (LSS) and the clustering properties of matter at very large scale can be scrutinized. Predictions of single-field inflaton models with a feature are already in disagreement with the data, Fig.[1-4], (also see \cite{cross}). 

Meanwhile there is no physical reason to expect that some of the couplings of various moduli fields, predicted generically by string theory, to the matter sector are highly suppressed.These coupling can for example give rise to a modified Friedman equation either on the geometry or the energy density side, in higher dimensional worlds and/or provide new noninflationary sources for LSS.

 Generically the coupling constants are functions of the $vev's$ of moduli fields $s$ and for this reason they can vary in space and time\cite{dvalikofman}. Therefore any slight or long-range variations in the string coupling constants to the matter sector or gravity, 
 while suppressing the autocorrelation functions $C_l^{TT}$ could also have a considerable impact on LSS and clustering properties at present horizon size scales. But $C_l^{TL}$ and $C_l^{TE}$ correlations compare the sources that give rise to $T,E$ spectra and the lensing potential of LSS thereby unravelling any discrepancies with inflationary physics.In short, unlike the single field inflation, it is natural for string theoretical models of cosmology to provide more degrees of freedom and more than one source for seeding LSS and $T,L,E$ spectra.

 We do not make use of the B-mode spectrum since measurement of the B-mode may not become available in a near future, although this spectrum, when available, can provide an independent handle for discriminating models through the consistency relation of the ratio of tensor to scalar amplitudes $r = A_t^{2} / A_S^{2}$ and cross-correlations with the $E, T$-modes.

Let us treat these cases one by one below. Our analysis is carried out with the data already available and the one expected in a near future from weak lensing measurements for large scales.

\subsubsection{ {\bf A.} { \it Inflaton potential bumps around $k_{60}$ }} 

These models were studied in \cite{easther} by introducing a step in the inflaton potential $V(\phi)$
\begin{equation}
V(\phi) = \frac{1}{2}m^2 \phi^2 \left[1 + c\tanh(\frac{\phi - \phi_{step} }{d})\right] \ ,
\end{equation}

This potential has a step at $\phi = \phi_{step}$ with size and gradient controlled by the two parameters $c,d$ respectively. The step can be chosen to correspond to $k_{60}$. It was shown in\cite{easther} that besides departure from scale invariance and an oscillatory behaviour of the CMB spectrum this feature can produce significant effects at large scales. Thus using the observational input one can in principle constrain the parameters $\phi_{step},c,d$ such that this model comes in agreement with CMB at all scales, including the suppression of power at low 'l'.

The extreme case of the potential step function is to take a complete cuttoff at $\phi_{step}$ thereby limiting the inflationary window to no more than 60 efoldings. This case is treated in \cite{linde} and the authors showed explicitly how this model can suppress quadropole power completely, and be in perfect agreement with the WMAP observations for $C_l^{TT}$.

However, the analysis of the cross-correlations spectra $C_l^{TE,TL,EE}$ were not investigated for either model until now. As shown in Fig. [1,2,4] in Sect.3, all the auto and cross-correlation spectra for the class of inflationary models with a feature (dashed line graph) exhibit the same behaviour in the low 'l' regime since they are sourced by the same term and thus highly correlated \cite{ma}.

The fact that all spectra and LSS are seeded by the same source, namely the primordial power spectrum, is a prediction of inflation. Therefore the modifications in spectra are overconstrained and do not allow any possibilities to simultaneously fit all data by e.g.simply introducing features on inflaton potentials. This means that if a feature on inflaton potential suppresses $C_l^{TT}$ at large scales than it will automatically imposed the same suppression on $C_l^{EE}, C_l^{LL}$ at large 'l'.

\subsubsection{{\bf B. } { \it String and Brane-World Models}}

There are many phenomenological models in literature for this class. Since we would like to make some generic and preferably model independent analysis and predictions, it is useful to try and classify them in as a general way as possible. For consistency, the first criteria to this purpose is to consider only models that may accommodate the observed late time acceleration of the universe thereby including the ISW effect on large scale spectra. We proceed in the classification in the following manner: first we classify them by the time when string corrections to conventional physics become significant; then independent of the time of deviation, we catalogue these models by all the possible imprints they can leave on the low energy observables.

{\bf B.1.} {\it Early Time String Models}. 

This is the group of models in class $B$ where string modifications to general relativity and quantum field theory are non-negligible in the early universe.

Friedman equation is modified by terms that grow with the energy density $\rho$, like
$[\rho / \sigma]^n$ where $\sigma$ is the brane tension taken to be of order GUT scale,
i.e. of roughly the same order as the height of the inflaton potential.
\begin{equation}
\label{early}
H^2 = G_N \left[ \rho +/- \rho (\rho /\sigma)^n \right] =\tilde G_N  \left [ \rho \right]   \ ,
\end{equation}

with 
\begin{equation}
\tilde G_N \simeq G_N \left[ 1 +/- (H_i^{1/2} / \sigma)^{n}\right ] \ ,
\end{equation}

$H_i$ denotes the Hubble parameter during inflation and $n \geq 1$ is some parameter of the theory depending on the specifics of the brane-world model. The sign depends on power $n$ and whether the brane tension $\sigma$ is positive or negative. Since $\rho$ dilutes with time while the string scale $\sigma$ is a fixed parameter, then the linear term dominates over the higher order correction terms at late time thereby recovering the conventional Friedman equation.

The case of $n=1$ recovers the Randall-Sundrum scenario\cite{rs}. The CMB spectrum suppression at large scales for this model was treated in detail in  \cite{katie} and it was shown that for a negative tension brane one recovers the power suppression observed by WMAP. Interestingly enough, by expecting the string description to be valid above and up to the inflaton potential height ($\sigma \simeq V_{max}$ ), the CMB coincidence puzzle is realized naturally in this model.

The string scale in these scenarios is of order $\sigma$. The ${TT}$ spectrum is similar to the class ${\bf A}$ models above, since the modified Friedman equation changes the primordial spectrum. But cross-correlation spectra, $C_l^{TE,TL,EE}$, can be different from class ${\bf A}$ by accounting for modified sources arising from moduli couplings to matter, (discussed below), Fig.[6].

{\bf B.2.} {\it Late Time String String Models}.  

String modifications to general relativity and/or quantum field theory become non-negligible in the late-time universe.

Friedman equation is modified differently such that the string correction terms  are
important at late times or low energies and not at high energies of order
the string scale $\sigma$.

Cardassian Model\cite{cardassian}, the DGP gravity \cite{dgp}, Kogan et al. or Gregory et al. models \cite{dgp,kogan,gregory} and the group of models described in \cite{dvaliturner},\cite{selftune,stoicatye}, are some representative examples for this class of late-time string models.

Although the specific physical scenario that gives rise to the modifications is different in each one of them, (some of them modify the LHS of Einstein equations, namely the gravity side, and some other models modify the RHS, namely the stress-energy tensor), they have one thing in common: they 'weaken' the strength of gravity' at late times/large distances and thus give rise to the late time accelerated expansion by various mechanisms, but without adding dark energy. Parametrically this group can be represented by the following modified equations

\begin{equation}
\label{late}
H^2 = \frac{8\pi G_N}{3} \left[ \rho + \frac{(1 - \frac{\rho}{\rho_c})H^{\alpha} }{H_0^{\alpha-2} }\right ] = \frac{8\pi \tilde G_N(z)}{3}\left [\rho \right] \ ,
\end{equation}

$\alpha$ is a parameter less than one and $H_0 \simeq 1/r_c$ is the present Hubble scale
given in terms of a crossover scale $r_c$ resulting from the higher dimensional nature of gravity in this string theoretical framework.

Formally one can attribute the correction term, obtained after reducing to a $4-dim$ low energy world, to a rescaling modification of Newton's constant $G_N$
\begin{equation}
\label{eq:gn}
\tilde G_N= G_N \left [1 + (\frac{r}{r_c})^{(1-\alpha)} \right ] \,
\end{equation}

with $r$ comoving physical distance by expressing the energy density $\rho(z)$ as a function of $r(z)$.

%%%%%%%%

Both groups of string theoretical models above as well as other string-inspired models that are not included in this analysis (e.g. \cite{selftune,stoicatye} ), can give rise to a set of imprints in the low energy observables which we catalogue below.

\subsubsection{ Possible String Imprints on a Low Energy World}

   {\bf i}) {\it Coupling of moduli to gravity/metric results in a modified Friedman equation}, at early or late times depending on the setup of the model, (e.g. \cite{cardassian,dgp,kogan,gregory,selftune,katie}). This case was treated above for the two groups of early and late time models for the specific examples of radion and dilaton couplings.

  {\bf ii}) {\it Coupling of moduli to matter sector gives rise to short or long-range $5^{th}$ forces and nonlinear dispersion relations} in the matter sector.
Interaction of the moduli fields, $s$, with the matter sector, predicted generically by string theory, will give rise to $5^{th}$ forces, which are highly constrained by experiments. Let us investigate this imprint more carefully below for both groups of models, {\bf B.1} and {\bf B.2}. The $5^{th}$ force type of interactions are subjected to severe constraints but its carrier  can contribute to stringy modification of clustering properties.
%%%%%%%%%%%%%%%%%%%%%%%55

{\it Early times models,} {\bf B.1.}: Any perturbations generated by moduli should obey the perturbation constraint $\frac{<\delta s>}{ < s >} \leq 10^{-4}$. For light moduli fields the fluctuations are of the order the horizon size, therefore the $vev$ of the field becomes of order

\begin{equation}
<\delta s> \simeq H_i \,\,\,  \,  <s> \simeq 10^4 H_i =O(\sigma) \ ,
\end{equation}
 The $<s>$ field would decay fast through coupling to other fields. Interaction of this moduli with other fields can be problematic since it may induce a large mass term on the other fields through radiative corrections. Quite likely couplings in the $TeV$ range may be forbidden by experiments since limits on the range of interaction $O(\sigma)$ are close to observational detection in collider experiments. BBN imposes another set of constraints on these short-range interactions with moduli. One can appeal to a moduli pseudoscalar or a global shift symmetry for $s$ in order to avoid such problematic couplings. The argument is similar to the treatment in \cite{5force} of constraints imposed on the quintessence interactions with the matter sector producing $5^{th}$ force effects.

Derivative couplings may not be suppressed by such symmetries. Coupling of moduli field $s$ to the EM stress tensor $F_{\mu\nu}$ is an example. It is given by the lagrangian

\begin{equation}
{\cal L}_{i} = g \frac{ s }{ M } F_{\mu\nu}{\tilde F^{\mu\nu}} \,
\end{equation}

where $M$ is the fundamental scale of the theory,$g$ is a coupling constant and $F_{\mu\nu}$ is electromagnetic field strength tensor (EM) and ${\tilde F}$ its dual. Similar couplings to QCD fields can be written by replacing the EM tensor $F_{\mu\nu}$ with the QCD field strength tensor $G_{\mu\nu}$. $M$ is the mass scale of the higher dimensional
theory,  that can be taken of order the $(3+1)$- brane tension $\sigma$, and $g$ is the coupling constant. Depending on the specific model, $\sigma$ can take some value anywhere from $M_{pl}$ or $GUT$ scale  up to $TeV$ scale.

For QCD fields this interaction would give rise to {\it short-range} forces of order $\sigma$ by breaking the shift symmetry. However the moduli field $s$ can decay into lighter particles in a relatively short time $\tau \simeq \sigma^{-1}$. Detection of these interactions may be an intriguing possibility in upcoming collider experiments.
 The situation is different for EM interactions, for example the background photons, since we can do an integration by part and express ${\cal L}_i$ as a derivative coupling which is not suppressed by the symmetry
\begin{equation}
{\cal L}_{i} = g ( \frac{s}{M} ) F^2 = \frac{g}{M} (\nabla_{\mu} s ) j^{\mu} \ ,
\end{equation}

with $j_{\mu}$ obtained after integrating by parts the EM tensor $F_{\mu\nu} = [D_{\mu} A_{\nu}]$.

So EM fields can have derivative coupling through the term $\frac{<\dot s>}{M}$. This interaction gives rise to a dispersion relation for photons
\begin{equation}
\label{dispersion}
\omega^2 = k^2 + k \frac{g (<\dot s>)}{M} \ ,
\end{equation}

The induced dispersion relation for photons may be reflected in a shift in the polarization plane of photons which can affect the $C_l^{TE}$ correlation. Similar couplings to electrons may also induce fluctuations. Depending on the energy scales of the specific string model, if these couplings are relevant during recombination epoch, then they also provide a mechanism for inducing fluctuations in the electron number $n_e$, that enter the expression for Thomson scattering, as well as fluctuations in the optical depth $\tau$ due to the variation of the string coupling constant $g$ (in general a function of $<\dot s>, s, \sigma$ ).These fluctuations depend on $< \dot s >$ as follows

\begin{equation}
x =  \frac{\delta \tau}{\tau}    \simeq    \frac{\delta n_e}{n_e} 
\simeq \delta [\log(<\dot s>)] \,
\end{equation}
where we take $g \simeq O(1)$, and $x \simeq \frac{\dot H_i}{H M}$

Early time models can give a hard cuttoff for all long wavelengths with  $k \leq x$ since ${<\dot s>}$
 is negative, ( due to $H_i$ slowly decreasing with time), in the dispersion relation for
photons. However $x$ is very small and this effect may be negligible. Besides, fluctuations in the optical depth alone are not sufficient for reaching agrement with data.

%%%%%%%%%%%%%%%%%%%%%%%%%%%%%%%%%%%%%%%%%%%%%%%%%%%%%%%5

{\it Late-time models} {\bf B.2.}: The interaction lagrangian to lowest non-renormalizable order ${\cal L}_{i}$ for coupling of the moduli field $s$ can be formally written here in the same way as for case {\bf B.1}. 

However the order of magnitude estimate for the range of values for $<\delta s>, <s>$ in this class are very different as seen in Eqn.(8,10). The $s$-field  for case {\bf B.2} has a range
of force of order the crossover scale $r_c$ taken to be roughly our present Hubble radius $r_c=H_o^{-1}$. Therefore coupling to these moduli gives rise to {\it long-range} $5^{th}$ forces. 
 Due to the severe constraints on the long-range 5th force experiments and variation
of the $\alpha$ constant, coupling to the QCD fields are forbidden in this group or have to be very
highly suppressed, \cite{5force}.

However derivative couplings to the EM field is not suppressed by the shift symmetry

\begin{equation}
{\cal L}_{i} = g \left ( \frac{ \nabla_{\mu} s }{M} ) \right) j^{\mu} \ ,
\end{equation}

Following the same steps as in case {\bf B.2}, ${\cal L}_{i}$ gives rise to a dispersion relation for photons

\begin{equation}
\omega^2 =k^2 + k \frac{g}{M}  (<\dot s>) \ ,
\end{equation}

Here $<\dot s> $ may be a positive quantity since the Hubble parameter $\dot H_0 \geq 0$ at low redshifts due to the accelerated expansion of the universe. 

The nonlinear behavior of the dispersion relation becomes important when the second term is dominant, $k < \frac{g <\dot s>}{M} \simeq g /[1+z]$ since roughly 
\begin{equation}
\label{sflux}
<\dot s> \simeq  \frac{H M}{[1+z]} \,\, < s > \simeq M \log[1+z]
\end{equation} 

The shift in the angle of polarization plane for photons is e.g. $\delta \theta \simeq (g/[1+z])$ where z is the
redshift (see also \cite{balaji}). The optical depth variations for these scenarios are very small, with an order of magnitude 

\begin{equation}
\frac{\delta \tau}{\tau} \simeq \delta (\log [H_{equality}]) \ ,
\end{equation}

where
equality means the mater radiation equality time, at low redshifts. This estimate follows from the relation $\delta [\log(<\dot s>] \simeq \delta \log H(z_e)$. The latter is a consequence  of energy conservation equation.

%%%%%%%%%%%%%%%%%%%%%%%%5

{\bf iii)} {\it Modulated string perturbations}. String coupling constants, $g$ are functions of moduli fields. Varying moduli vev's give rise to varying coupling constants which in turn can generate a new channel of noninflationary perturbations, example \cite{dvalikofman}

\begin{equation}
\label{varg}
\frac{\delta g(s)}{g(s)} \simeq \frac{\delta T}{T} \simeq \frac{\delta \Phi[r,z]}{\Phi[r,z]}
\end{equation}

 This effect of modulated perturbations was treated in \cite{dvalikofman}where a suppressed $B$-mode was offered as a signature on the spectra from the varying of coupling constants. Since the B-mode data is not available, cross-correlations can be used to identify the imprints of varying string couplings on the source term from the data available, as discussed in Sect.3.

{\bf iv)} {\it Different clustering at large scales}. Any of the string imprints in $(i-iii)$ above can change the clustering properties of large scales by modifying the source term of perturbations or the background gravitational potential $\Phi$ for perturbations, e.g Eqn.~(\ref{varg}). Since moduli coupling to matter sector and varying coupling constant contribute with a new channel of perturbations to the spectra besides the inflationary ones then the gravitational background potential receives corrections of the form

\begin{equation}
\Phi \simeq \Phi_{inflation} + \phi \left [g, \frac{<\delta s>}{<s>}\right] \ ,
\end{equation}
 The new contribution to the source term in the spectra and the correction to the Newtonian potential of the background, $\phi [ g, \frac{<\delta s>}{<s>} ]$, if significant at late times, can change the clustering properties at very large scales. The details of $\phi [g, \frac{<\delta s>}{s}]$ depend on the specific string model considered. For example this correction term is negative for DGP type models as given in \cite{dvalistarkman}. Another example is when moduli coupling to relativistic matter gives rise to long range $5^{th}$ forces in which case the carrier of the force introduces a correction $\phi [g, \frac{<\delta s>}{s}]$ to the gravitational potential that, for a coupling constant $g$ proportional to some power $(1- \alpha)$ of moduli, Eqn. (\ref{sflux}), in case {\bf B.2}, becomes

\begin{equation}
\label{phi}
\frac{\delta g}{g} \simeq \frac{<\delta s>}{<s>} \simeq \frac{\delta \Phi}{\Phi}=\frac{\phi}{\Phi}=\left(\frac{r[z]}{r_c}\right)^{1-\alpha}
\end{equation}

where  $r_c$ denotes the combination $r_c = g/M $.

The correction term can be identified and deduced from $C_l^{TL}$, Eqn.~(\ref{temp1} - \ref{lens2}). If the expression for $\phi[g, \frac{<\delta s>}{<s>}]$ contains a new string scale then we could deduce the emergence of this new stringy scale related to the coupling constant $g$.  A coupling constant that varies slowly over cosmological scales, (e.g. oscillating every one Hubble time), is such an example of a significant contribution to clustering at large distances. See Fig.[4,5].

Since a feature in the inflaton potential will not carry over to LSS, and can not give rise to more than one source term in the perturbation spectra (unlike the extra degrees of freedom provided by string imprints in $(i-iv)$) , then a consistency check from cross-correlating with the $E-$mode and cosmic shear $L$ reveals the stringy modifications to the sources Fig.[1-7], thus providing us with information on the true origin of $C_l^{TT}$ power suppression .

With this treatment and classification of stringy effects on low energy observables, we are now ready to discriminate imprints of inflaton features from the stringy signatures on the origin of the observed anomalies in the spectra in Sect.3 through the comparison of spectra $C_l^{TT,TE,EE,TL}$.

This analysis can also be used to discriminate among models within Class {\bf B}, since while class {\bf B.1} of Early Times modifications distorts the LSS and CMB spectrum through modifications of the primordial spectrum, the class {\bf B.2.} of Late Times modification will manifest its signature as a modification to the ISW effect as well.

%%%%%%%%%%%%%%%%%%%%%%%%%%%%%%%%%%%%%%%%%%%%%%%%%%%%%%%%%%%%%%%%%%%%

%%%%%%%%%%%%%%%%%%%%%%%%%%%%%%%%%%%%%%%%%%%%%%%%%%%%%%%%%%%%%%%%%%%%

\section{Analysis of present data}

\subsubsection{Data Analysis}

First let us bear in mind that most of the string and brane-world models are phenomenological. A rigorous derivation of a realistic cosmology is yet to be obtained from string theory. As such these models share a set of problems and open questions related to the lack of  a consistent effective theory that would replace general relativity. Lacking the underlying theory, means that in most cases we do not have the calculational framework for predicting the perturbation equations and thus face the following challenges: perturbation equations may be different from the standard ones \cite{Deffayet:2001pu,Deffayet:2002sp}; Poisson equations and Boltzmann transport equation for the cosmic fluid of perturbations may be different for the modified theories \cite{gondolo}; Einstein equations may be all modified but only the stringy  modifications to the Friedman equation are known while the corrections to the other components of $G_{\mu\nu}$ and a generalized Bianchi identity are still unknown. For these reasons a pedantic and generic calculation of their predicted CMB spectra and the hunting for their imprints on the spectra becomes difficult.

%%%%%

Nevertheless, with the information available and assumptions made about perturbation equations in these models, we carry out the calculation of their spectra and try to narrow down which models fit the data best.

\subsubsection{Large Scale Structure (LSS).}

At present there are two large galaxy surveys of comparable size, the
Sloan Digital Sky Survey (SDSS) \cite{Tegmark:2003uf,Tegmark:2003ud}
and the 2dFGRS (2~degree Field Galaxy Redshift Survey) \cite{2dFGRS}.
Once the SDSS is completed in 2005 it will be significantly larger and
more accurate than the 2dFGRS. In the present analysis we use data from SDSS, but the results would be almost identical had we used 2dF data instead. In the data analysis we use only data points on scales larger than $k = 0.15 h$/Mpc in order to avoid problem with non-linearity.

\subsubsection{Cosmic Microwave Background.}

The temperature fluctuations Eqn.(1) are conveniently described in terms of
the spherical harmonics power spectrum $C_l^{TT} \equiv \langle
|a_{lm}|^2 \rangle$, where $\frac{\Delta T}{T} (\theta,\phi) =
\sum_{lm} a_{lm}Y_{lm}(\theta,\phi)$.  Since Thomson scattering
polarizes light, there are also power spectra coming from the
polarization. The polarization can be divided into a curl-free $(E)$
and a curl $(B)$ component, yielding four independent power spectra:
$C_l^{TT}$, $C_l^{EE}$, $C_l^{BB}$, and the $T$-$E$ cross-correlation
$C_l^{TE}$.

The WMAP experiment has reported data only on $C_l^{TT}$ and $C_l^{TE}$
as described in
Refs.~\cite{Spergel:2003cb,Bennett:2003bz,%
Verde:2003ey,Peiris:2003ff}.  We have performed our
likelihood analysis using the prescription given by the WMAP
collaboration~\cite{wmap,Spergel:2003cb,%
Verde:2003ey,Peiris:2003ff} which includes the
correlation between different $C_l$'s. Foreground contamination has
already been subtracted from their published data.

\subsubsection{Likelihood analysis}

For calculating the theoretical CMB and matter power spectra we
use the publicly available CMBFAST package. As the
set of cosmological parameters other than those related to modified gravity we choose the minimum standard model with 6
parameters: $\Omega_m$, the matter density, the curvature
parameter, $\Omega_b$, the baryon density, $H_0$, the Hubble
parameter, and $\tau$, the optical depth to reionization. The normalization of both CMB and LSS spectra are taken to be free and unrelated parameters.

We restrict the analysis to geometrically flat models $\Omega = 1$.

\begin{table}
%\begin{ruledtabular}
\begin{center}
\begin{tabular}{lc}
%\colrule
\hline
parameter & prior\cr
%\colrule
\hline
$\Omega_m$ & $0-1$ (Top hat) \cr
$h$ & $0.5-1.0$ (Top hat) \cr
$\Omega_b h^2$ & $0.014-0.040$ (Top hat) \cr
$\tau$ & $0-1$ (Top hat) \cr
$b$ & free \cr
\hline
\end{tabular}
\end{center}
%\end{ruledtabular}
\caption{The different priors on parameters
used in the likelihood analysis. Parameters related to modified gravity are not tabulated here.}
\label{table:prior}
\end{table}

Models with modified primordial spectra which suppress TT power at large scales have already been extensively analyzed in the literature, (Sect.2, class {\bf A, B.1}). Our analyses show that the $T, E$ low multipoles cannot simultaneously be fitted by a break in the primordial power spectrum. The main reason is that $TT$ and $TE$ power spectra are affected in the same direction by a change in the initial spectrum, Fig.[1-3] ({\it dashed-line}). On the other hand the WMAP shows evidence for low $TT$ power and high $TE$ power at small multipoles. Even adding a high optical depth the spectra cannot be fitted to observation.

We have performed a likelihood analysis of late-time models by using  Deffayet et al. \cite{Deffayet:2001pu,Deffayet:2002sp} as an illustration (see also \cite{cardassian} and \cite{dvaliturner})
where the Friedman equation is modified directly). In Deffayet et al. \cite{Deffayet:2001pu,Deffayet:2002sp} model there is no change in the perturbation source terms.

We take this model mainly as an illustrative example. Even though it provides a good fit to CMB and LSS data, it is a quite poor fit to the combination of all available data including measurements of the Hubble parameter. Parameter fitting to more general modifications of the Friedmann equation can for instance be found in \cite{elgaroy}

Furthermore, as discussed previously it is natural that a change in the effective strength of gravity also leads to modified source terms for the perturbation equations. In the likelihood analysis we therefore adopt the approach that besides the modifications in the Friedman equation , case {\bf B.2} \cite{cardassian, dvaliturner, Deffayet:2001pu}, the perturbation source is also modified according to $G_N$ in Eq.~(\ref{eq:gn}) with $\alpha=0$. We call this case {\bf a}. 

To check the effect of just modifying the perturbation source terms we have also performed the likelihood analysis for the concordance $\Lambda$CDM model with modified perturbation source terms (but with no modification to the Friedman equation), again modified as in Eq.10) with $\alpha=0$. We call this case {\bf b}.

In both cases we take $r_c$ as a free parameter in the fit. In order for the Deffayet et al. model to be self consistent, $r_c$ is already fixed by the modified Friedman equation. However, since both the modified Friedman equation and the modified $G_N$ are effective equations there is no fundamental reason why they should be modified in the exactly the same way since in a more general theoretical framework both modifications can occur independently.

\begin{figure}
\includegraphics[width=80mm]{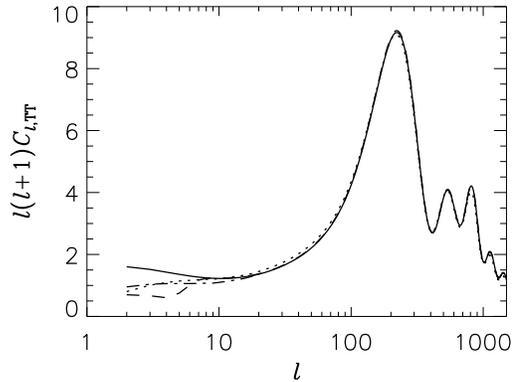}
\caption{$TT$ spectra of the various models. The full line is the standard $\Lambda$CDM model, the dashed line has a modified primordial power spectrum, the dot-dashed a modified perturbation source term (case {\bf b}). Finally, the dotted line has a modified Friedman equation and modified source terms (case {\bf a}). Normalization is arbitrary.}
\label{fig:cl}
\end{figure}

In Fig.~\ref{fig:cl} we show the best fit models for cases {\bf a} and {\bf b}. Both have $k_c = 2\pi/r_c \simeq  3 \times 10^{-6} \,\, h$/Mpc.
In addition we show the standard $\Lambda$CDM model, and inflationary models with 
broken scale invariance. This model is a concordance $\Lambda$CDM model with no power at $k < 5 \times 10^{-4} \,\, h$/Mpc, the same model studied in Ref.~\cite{kesden}.

On small scales all the models are clearly indistinguishable, but at low multipoles there are significant differences.

\begin{figure}
\includegraphics[width=80mm]{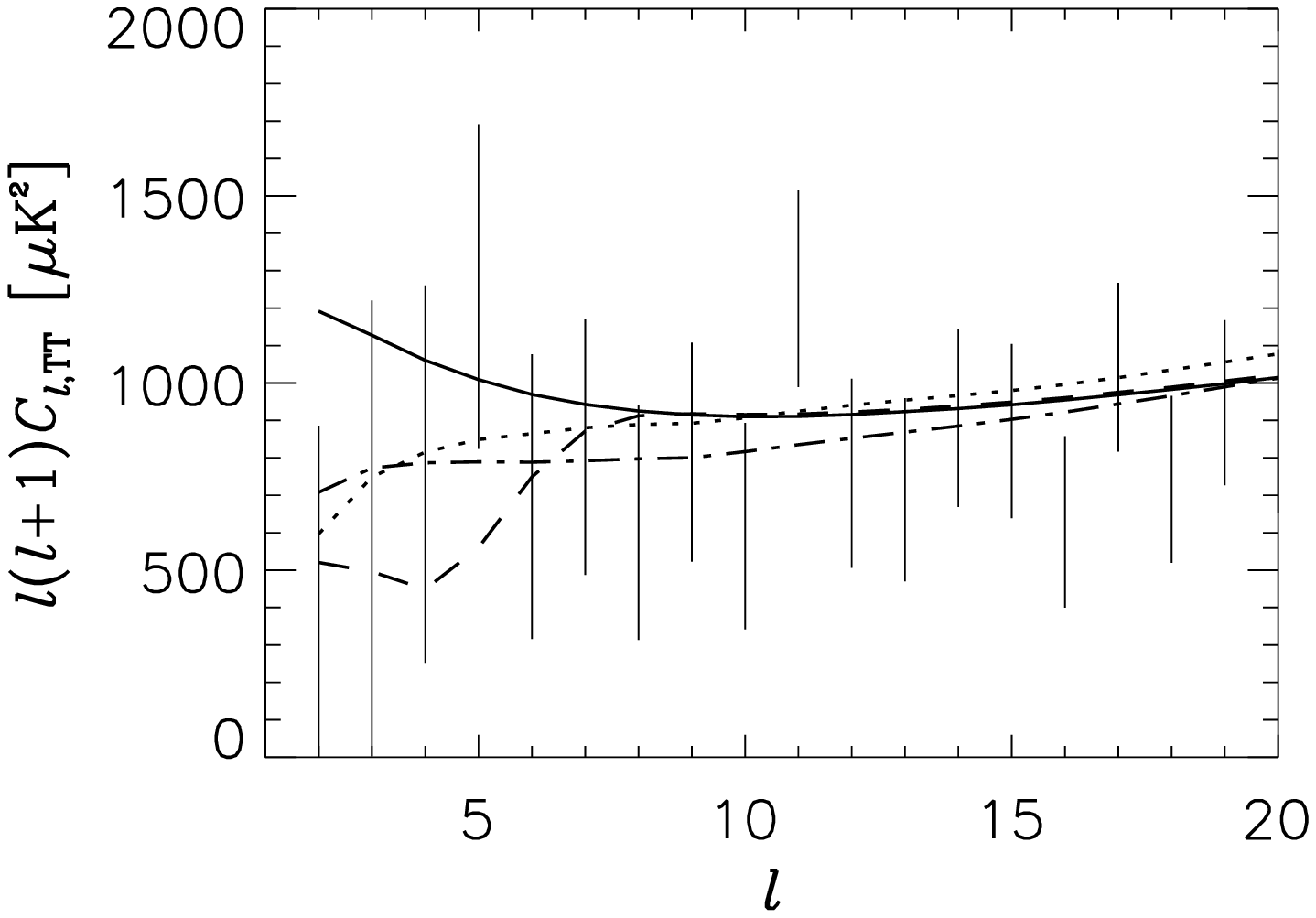}\\
\includegraphics[width=80mm]{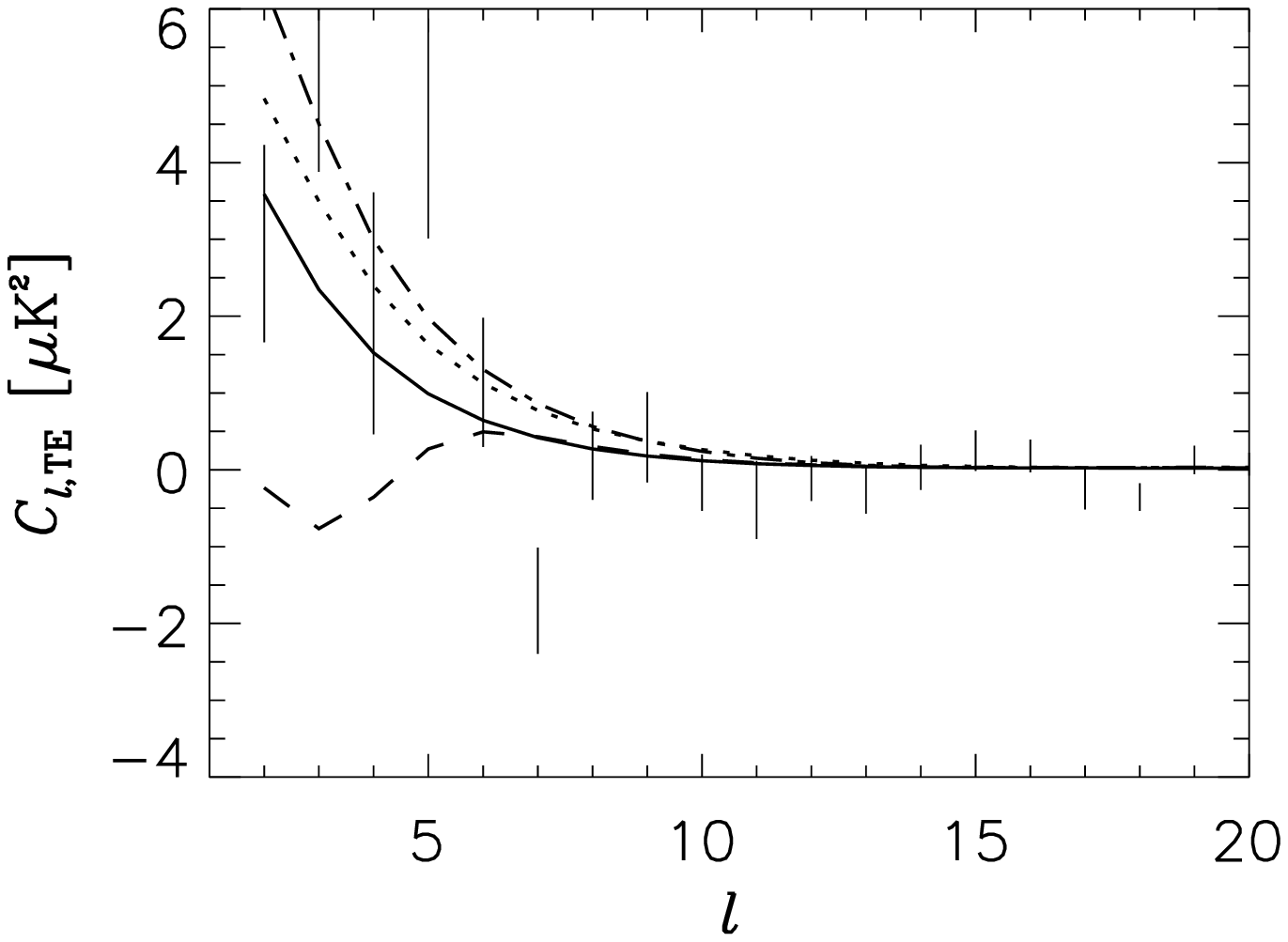}
\caption{$TT$ and $TE$ spectra of the various models. The full line is the standard $\Lambda$CDM model, the dashed line has a modified primordial power spectrum, the dot-dashed modified perturbation source terms (case {\bf a}). Finally, the dotted line has a modified Friedman equation and modified source terms (case {\bf b}). The data points are those measured by WMAP}.
\label{fig:cl2}
\end{figure}

We show the large scale spectra in Fig.~\ref{fig:cl2}, both for $TT$ and $TE$. From this it is clear that the models with modified gravity are able to simultaneously produce less $TT$ power and more $TE$ power, and are in slightly better agreement with data than the standard model. The model with broken scale invariance is clearly a quite poor fit.

\begin{figure}
\includegraphics[width=80mm]{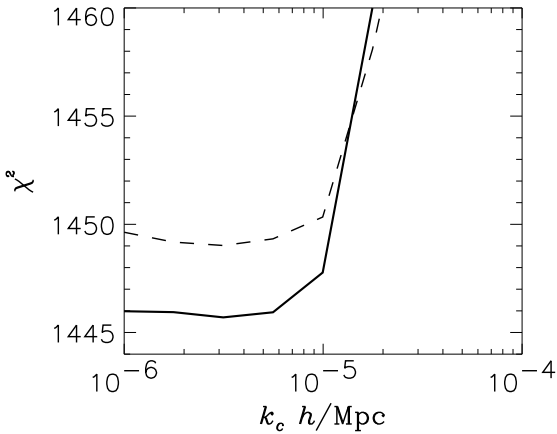}
\caption{$\chi^2$ as a function of $k_c$. The full line is for case {\bf a} and the dashed for case {\bf b}.}
\label{fig:chi2}
\end{figure}

In Fig.~\ref{fig:chi2} we show $\chi^2$ as a function of $k_c$ for the two cases. From that it is clear that models with modified perturbation source terms can be slightly better fits than models with standard perturbation equations, even though the effect is quite small.

However, even though case {\bf a} provides a better fit to CMB and LSS data, it does do for non-standard values of other cosmological parameters, such as $h$ and $\Omega_m$. These values are in  disagreement with other cosmological measurements, such as the HST key project measurement of $h$ \cite{freedman}

One should also note that if $r_c$ is taken to be fixed by the modified Friedman equation then, $k_c = 2\pi/r_c \simeq 10^{-3}$ and the $\chi^2$ of the best fit model is roughly $\chi^2=1.6 \times 10^4$, i.e.\ it is completely ruled out.

\begin{table}
\begin{center}
\begin{tabular}{lccc}
\hline Model & $\chi^2$ & $\Omega_m$ & h \cr \hline 

$\Lambda$CDM & 1449.5 & 0.30 & 0.69 \\

Case B & 1449.0 & 0.31 & 0.69 \\

Case A & 1445.7 & 0.42 & 0.58 \\
\hline

\end{tabular}
\end{center}
\caption{Best fit $\chi^2$ for the standard 6 parameter $\Lambda$CDM model, as well as modified gravity cases {\bf a} and {\bf b}. We also show the best fit values of $\Omega_m$ and $h$.} \label{table:parameters}
\end{table}

\subsection{Direct modification of the graviational potential}

\begin{figure}
\includegraphics[width=80mm]{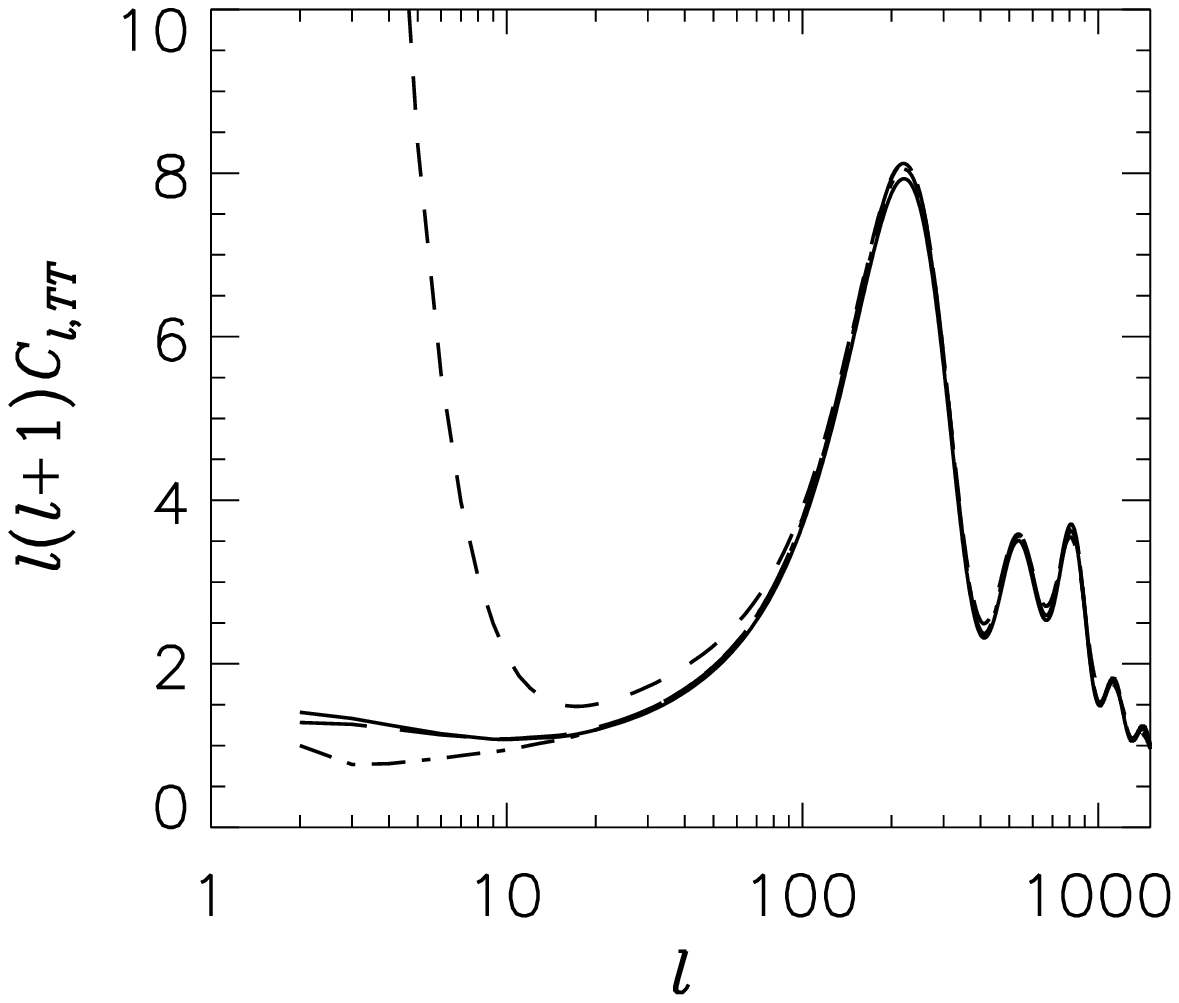}\\
\includegraphics[width=80mm]{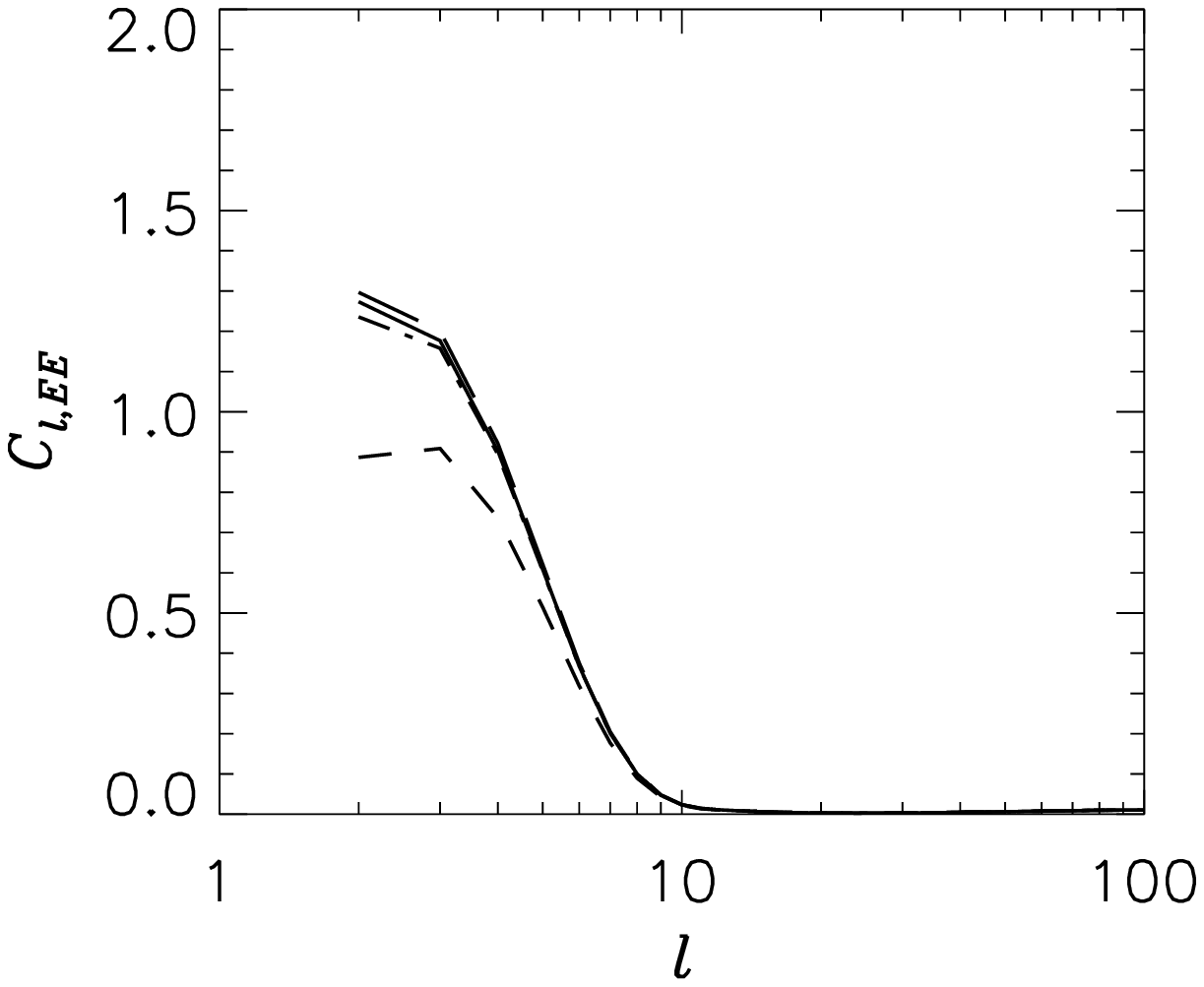}
\caption{$TT$ and $EE$ spectra of models with a modified $\phi$ according to Eq.~(\ref{phi}). The curves are for $\alpha=0$ and various values of $k_c = 2\pi/r_c$. The full line is for $k_c=0$, the long-dashed for 
$k = 10^{-6} \,\, h$/Mpc, the dot-dashed for 
$k = 10^{-5} \,\, h$/Mpc, and the dashed for
$k = 10^{-4} \,\, h$/Mpc. Normalization is arbitrary.}
\label{fig:mod1}
\end{figure}

\begin{figure}
\includegraphics[width=80mm]{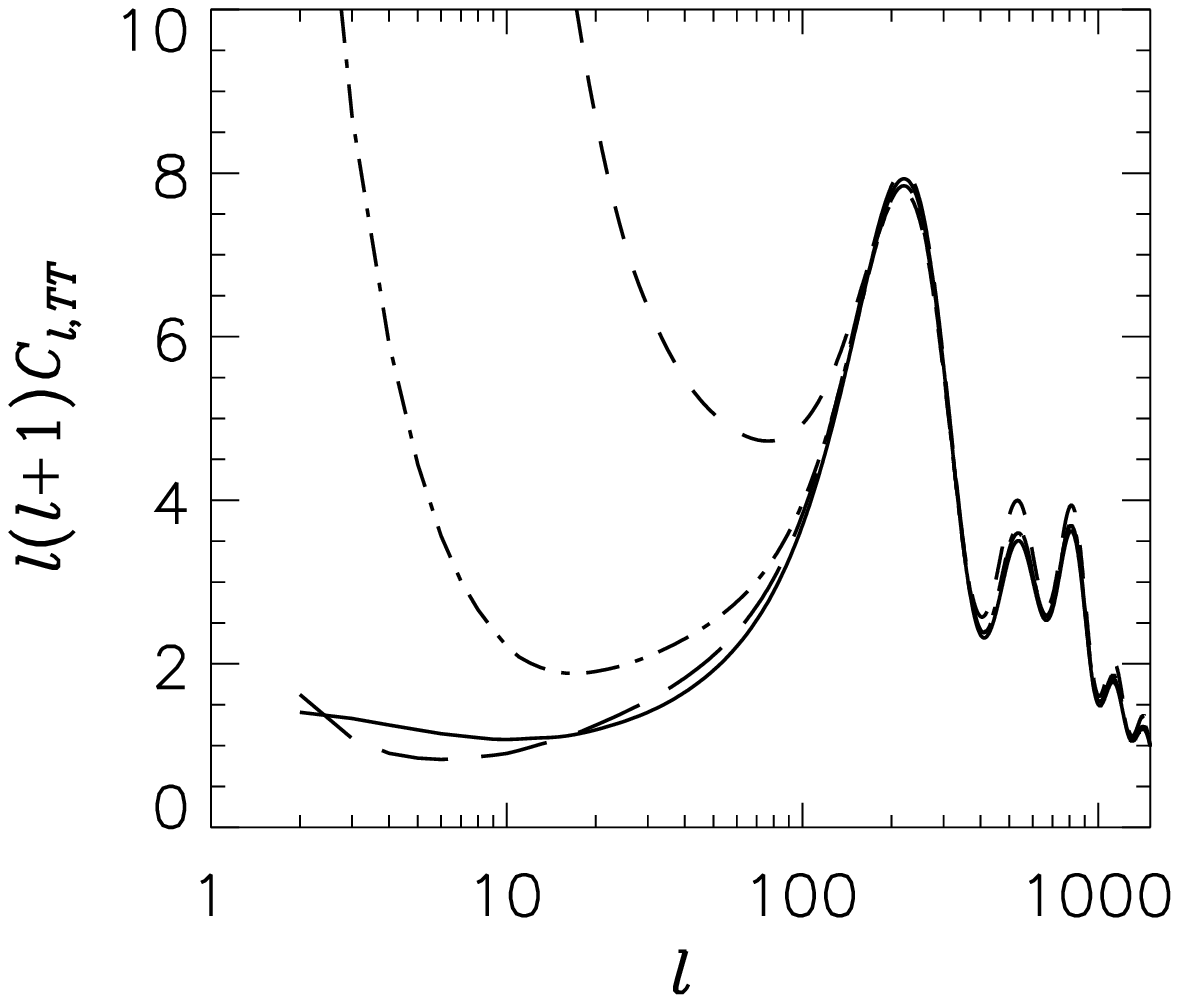}\\
\includegraphics[width=80mm]{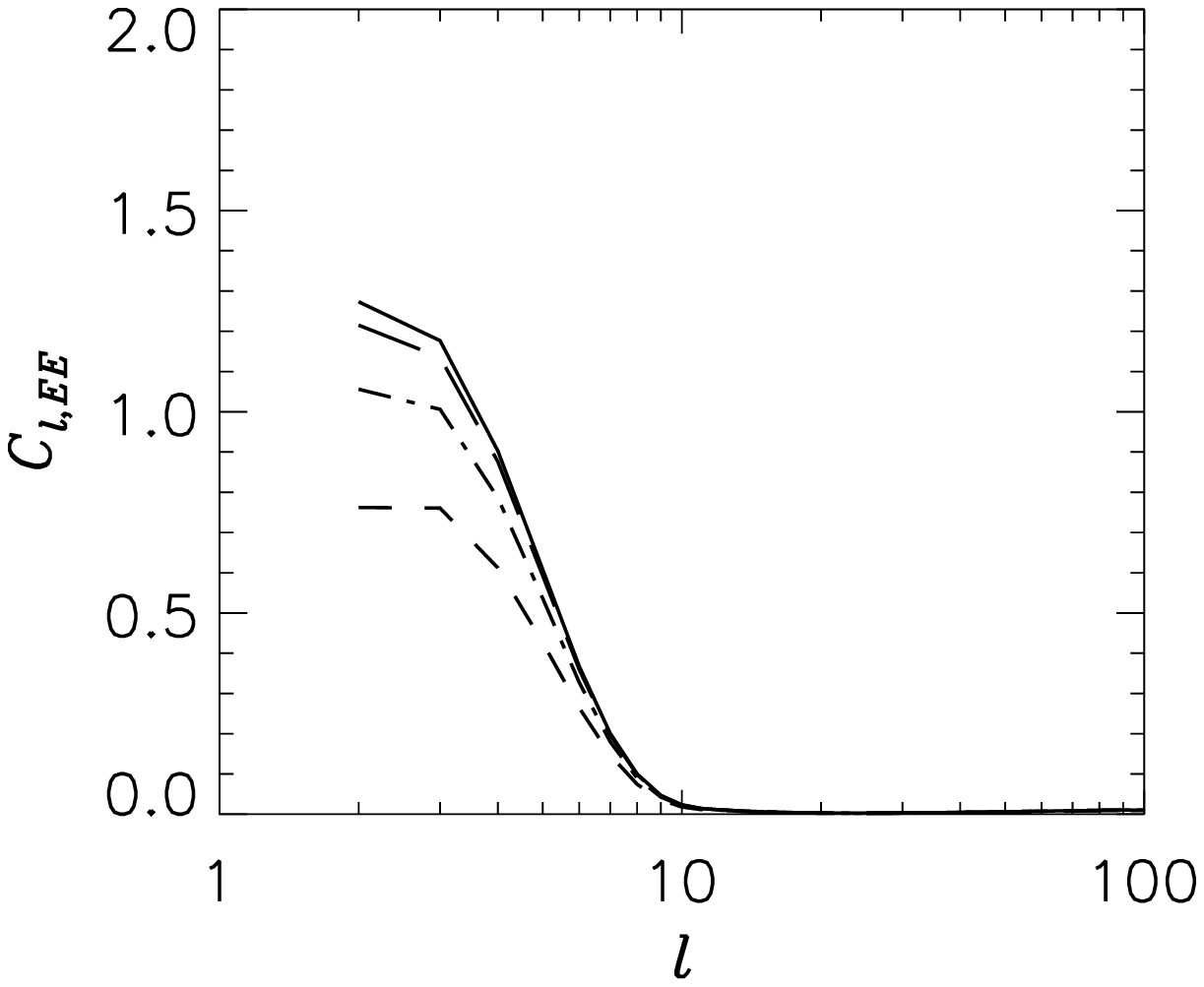}
\caption{$TT$ and $EE$ spectra of models with a modified $\phi$ according to Eq.~(\ref{phi}). The curves are for $\alpha=0.5$ and various values of $k_c = 2\pi/r_c$. The full line is for $k_c=0$, the long-dashed for 
$k = 10^{-6} \,\, h$/Mpc, the dot-dashed for 
$k = 10^{-5} \,\, h$/Mpc, and the dashed for
$k = 10^{-4} \,\, h$/Mpc. Normalization is arbitrary.}
\label{fig:mod2}
\end{figure}

Fig.[4,5] shows the analysis for the case of a direct modification to clustering properties at large scales by stringy corrections to the background gravitational potential. The specific case shown in the plot is the correction $\phi$ of Eqn.~(\ref{phi}) which, although of different origin, resembles the correction to Newton's constant of the DGP model and thus modifies the LSS source similarly. 

Interestingly enough, the turnover scale which is proportional to the string coupling constant $g(s)/M \simeq r_c$ has to be larger than the present Hubble radius to reach agreement with data.

By a similar analysis applied to the DGP model, the turnover scale of order-one modifications in $G_N$ has to be larger than $H_0$. The scale the authors of \cite{dgp} take $r_c \simeq  H_0$ in order to explain the acceleration of the universe is disfavored by data.

From Figures~\ref{fig:mod1} and \ref{fig:mod2} it can be seen that changes in the the temperature power spectrum are very similar to those obtained from modifying $G_N$, as could be expected. However, we also show the $E$-polarization power spectra, and interestingly there is almost no modification there. The reason is that the polarization anisotropy is solely related to physics around the epoch of recombination and to possible reionization. However, only the most extreme of the models shown have any significant modification to $\phi$ around recombination (as can be seen from the $TT$ spectra).

\begin{figure}
\includegraphics[width=80mm]{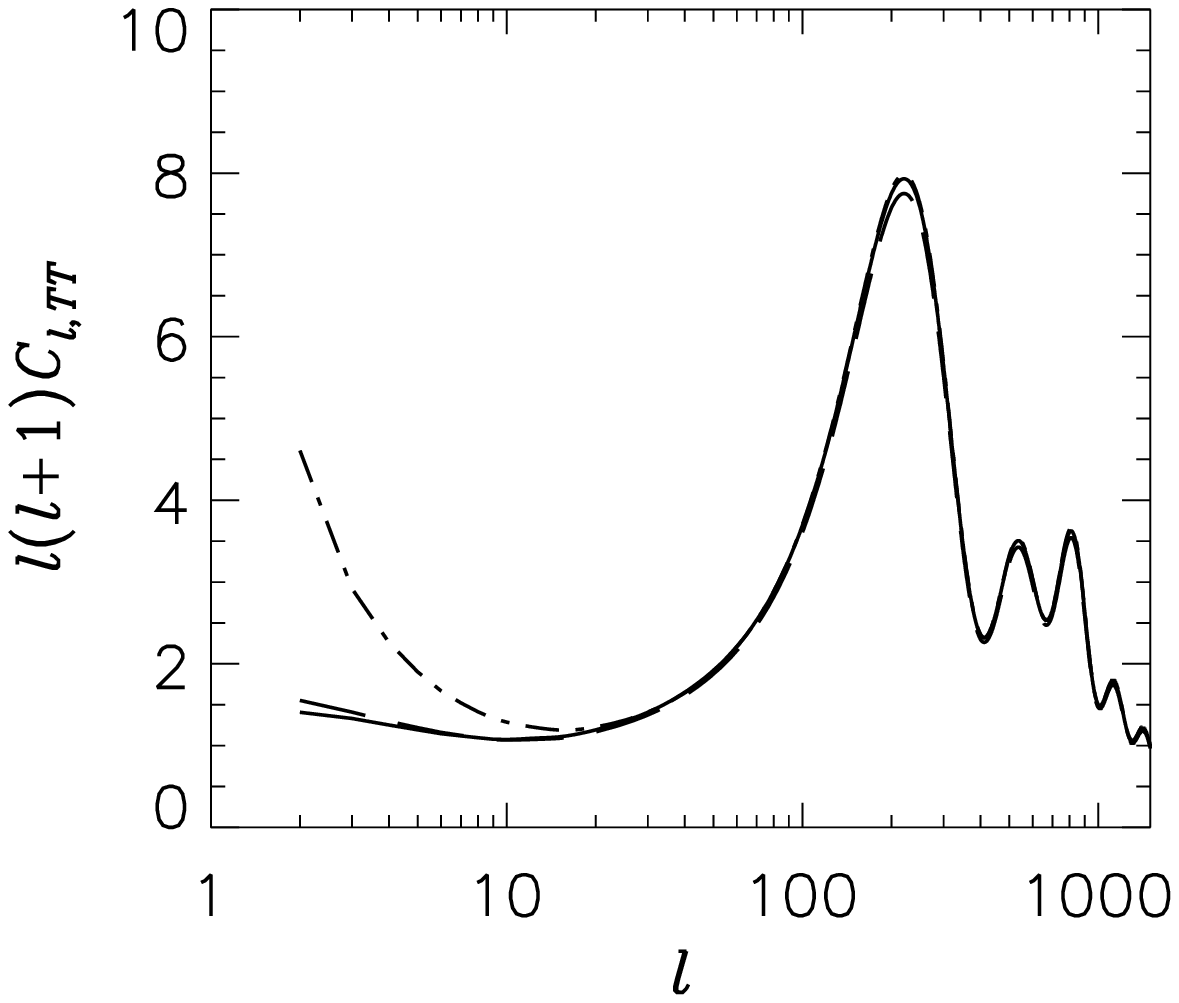}\\
\includegraphics[width=80mm]{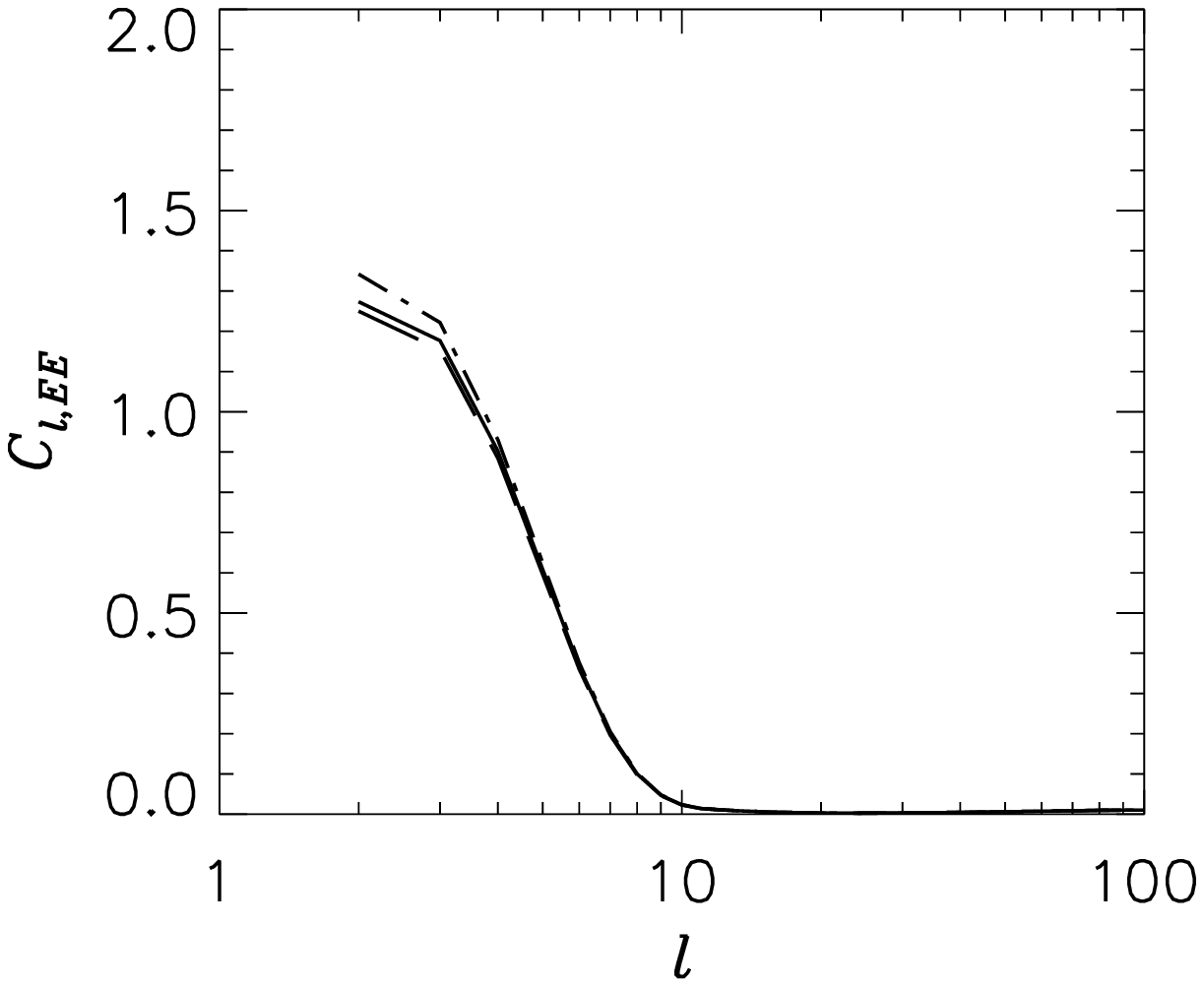}
\caption{$TT$ and $EE$ spectra of models with a modified $\phi$ according to Eq.~(\ref{phi}), but with NEGATIVE sign. The curves are for $\alpha=0$ and various values of $k_c = 2\pi/r_c$. The full line is for $k_c=0$, the long-dashed for 
$k = 10^{-6} \,\, h$/Mpc, and the dot-dashed for 
$k = 10^{-5} \,\, h$/Mpc. Normalization is arbitrary.}
\label{fig:mod3}
\end{figure}

In Figure~\ref{fig:mod3} we show spectra for direct modification of $\phi$, but for the case of a negative sign in Eqn.~(\ref{phi}) that can arise from a deconfining $5^{th}$ force. From this figure it is clear that there is no power suppression at low $l$ for the $TT$ spectrum. In turn that means that models which reduce the potential through negative contributions from the correction $\phi$ are poor fits to the observational data.

\subsubsection{Future data}

Since future observations will at most improve the very large scale precision moderately because it is already dominated by cosmic variance, it is an interesting question whether there is any other way to distinguish the various types of models.

\begin{figure}
\includegraphics[width=80mm]{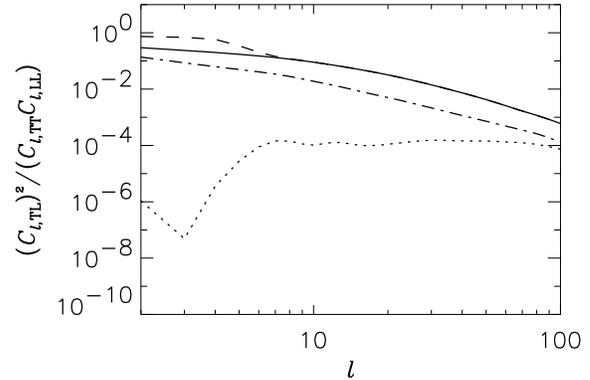}
\caption{Cross correlation between shear and temperature. The full line is the standard $\Lambda$CDM model, the dashed line has a modified primordial power spectrum, the dot-dashed modified perturbation source terms (case {\bf b}). Finally, the dotted line has a modified Friedman equation and modified source terms (case {\bf a}).}
\label{fig:cross}
\end{figure}

We follow the approach of Ref.~\cite{kesden} in looking at the large scale cross correlation between shear and temperature fluctuations, $C_l^{TL}$. Fig.~\ref{fig:cross} shows the amplitude of this cross correlation in the same way as in \cite{kesden}. In agreement with their result we find that the modified power spectrum model produces increased correlation at large scales. Very interestingly we also find that in case {\bf a} (modified Friedman equation and source terms) there is essentially no correlation between temperature and shear. That this is a feature of the modified Friedman equation can be seen from the fact that the correlation for case {\bf b} (only modified source terms) resembles the standard model.
This difference should be clearly visible in future surveys which would find no detectable $TL$ power for the modified Friedman equation models.

With respect to the upcoming $E$-polarization measurements from WMAP, the models with stringy modifications of gravity at late times predict little or no change in $E$-polarization compared with standard $\Lambda$CDM. This is certainly a testable prediction in the sense that early time modifications to the power spectrum directly influence all power spectra, including $C_l^{EE}$.

Finally, it should also be noted that we have not studied observational constraints on modifications of the photon dispersion relation, Eq.~(\ref{dispersion}). Effectively such a modification would lead to changes in both the power spectra and in the observed Planck spectrum of the CMB, similar to a chemical potential (see \cite{dispersion} for bounds on this possibility). Certainly this is something that merits further study, but is outside the scope of the present paper.We did not study the case of multifield inflation here.

\section{Do we have evidence for string theory?}

The analysis carried out in this paper of cross-correlations at large scales, of $T$ with polarization $E$ and lensing potential $L$, provides the first direct evidence for the existence of new physics.

The $C_l^{TE}, C_l^{TL}$ correlations can uniquely identify imprints of stringy modifications and discriminate these signatures from features on the inflaton potential by cross-examining the mismatch in the sources that seed the spectra. Comparison of $C_l^{TT}, C_l^{TE}, C_l^{TL}$ reveals that the sources which produce these spectra differ from one another. By a fit of real data to standard and string motivated cosmological models, it is clear that at least two different sources or equivalently two different scales are required in order to reach agreement with observation. One of these scales is related to the significance of the ISW effect on CMB produced by present acceleration and, the other one is related to $C_l^{TT}$ and modifications of horizon size clustering properties,i.e. at scales much larger than the recombination era in order to preserve the observed high $C_l^{TE}$.

\smallskip 

Models with features on inflationary potentials (Sect.2 case {\bf A}) or Early-Time brane-worlds (Sec.2 case {\bf B.1}) with modified Friedman equation which change {\it only} the primordial spectrum, but do not provide additional sources for modifying the late-time evolution and large scale clustering
in general are not able to explain the {\it high} $C_{TE}$ and the {\it low} $C_{TT}$ simultaneously and highly disfavored even with present data. The reason lies in the fact that, while it is possible to have additional sources in string-motivated models, inflationary cosmology can provide only one source for all spectra. Therefore modifying the primordial source by adding features to the inflaton potential, will simultaneously produce the same effect on all spectra $C^{TT}, C^{TE}, C^{TL}$, namely a suppression on $C^{TT}$ will also impose a suppression of $C^{TE}, C^{LL}$. Clearly this prediction of conventional single-field inflationary physics is already disfavored by WMAP data. Upcoming weak lensing surveys will provide further evidence through data for $C^{TL}$ in the existence of more degrees of freedom. If there is a cuttoff on the inflaton primordial spectrum then correlation $C^{TL}$ should be of order one and $C^{EE}$ spectrum highly suppressed at the cuttoff scale.

(Here we considered only single-field inflationanry models. Multi-field models involve Brans-Dicke models, coupling of the inflaton to some other field, like a Kaluza Klein mode for example or modifications to the Einstein action by higher derivative terms. These theories predict nonadiabaticity at large scales and production of isocurvature modes. Many of them are highly constrained by data. Besides, the input of new physics is involved in most of the above modifications).

String theory however can provide more than one source for the origin of perturbations and clustering at large scales. This fact contains the unique handle for our method of detection of imprints for string theory.
%%%%

New sources or modification to the background potential can arise, in string theoretical cosmology, by varying string coupling constants and moduli coupling to the matter sector, modified Friedman equation, noninflationary channel of perturbations, induced dispersions and very short or very long range $5^{th}$ forces,etc. in addition to the inflationary source of perturbations, (as described in Sect.2). These sources can generically provide noninflationary channels of perturbations, modified clustering or modified ISW effect. In short new physics allows for the existence of more than one source for $T,E,L$ and the emergence of a second scale, which as indicated by data fits the observation.

%%%

From our analysis of models with a standard $\Lambda$CDM cosmology, but modified gravitational potential according to Eq.~(\ref{phi})
it was clear that they are very similar to modifying $G_N$ on large scales. Such models are slightly better fits to the present data, if there is a modification to gravity as a scale of $k_c \sim 10^{-6}-10^{-5} \, h$/Mpc, which is significantly larger than the present Hubble horizon. We are just beginning to see a glimpse of string theory modifications to conventional cosmology. Weak lensing for large scales becomes an important direction for precision cosmology as it may provide further important evidence for new physics.

%%%%%%%%%%%%%%%

It is interesting that the scale for late-time modification of gravity needs to be much larger than what is found in the DGP model. This model in its original form is clearly ruled out by present cosmological data. However the need for $r_c \gg H_0$ seems to be true in general for class {\bf B.2} of Sec.2 and, can be interpreted as evidence for the emergence of a second scale in physics, which is not mediated by inflation.

%%%%
It should be noted that none of the popular stringy models we analyzed here, Fig[1-7], are a perfect match to data. However this does not affect the conclusions in this work. Our scope here was not to advocate or rule out specific models but rather to make a generic model-independent prediction whether evidence for new physics exists. Upcoming data from weak lensing surveys and data from WMAP will improve our bounds on modifications and allow an analysis of models on a case by case basis. With the current data available for the acceleration of the universe and the combined data of CMB+LSS our result is that the best fit are models that contain two modified sources: a late-time modified Friedman equation which uniquely alters the ISW effect; as well as a  modified perturbation source which changes the clustering properties at large scales.A generic prediction for this type of string modifications is the lack of correlations between $T,L$. If weak lensing data finds a suppressed $C^{TL}$ spectrum at large scales than this is strong evidence for the existence of two different sources for $T$ and $L$.

\smallskip

Future large scale cosmic shear measurements should be able to see a clear difference between the stringy models and the conventional $\Lambda$CDM models, with or without modified spectra.

%%%%

Also, the upcoming measurement of $E$-type polarization by WMAP may give clues to the physical nature of the large scale features. Models with late time modified gravity predict essentially no modification in the $E$-polarization spectrum, as opposed to models with early time gravity modifications and/or inflationary potential features which do predict a direct modification of the primordial power spectrum and therefore also of the $E$-polarization. Therefore improvement of $C^{TE}$ data expected in a near future from WMAP will allow us to better discriminate among various models and mechanisms.

\smallskip

'Reading' new physics in the sky makes for  extremely exciting times. Currently there are interesting anomalies in the CMB at the very largest scales. At present there are many models in literature which aim at explaining them. Models with modified gravity on very large scales and modified clustering look promising. After all it may not be a coincidence that the accceleration of the universe and CMB power suppression occur at the same scale.

With data available in the near future, such as polarization measurements of the CMB by WMAP, and weak lensing surveys, it will be possible to exclude many of the models and collect further evidence for new physics.{\it Cosmological observations thus provide the first window to string theory}.

\end{document}